\RequirePackage{fix-cm}
\documentclass{svjour3}                     
\smartqed  
\usepackage{graphicx}
\usepackage {threeparttable}
\usepackage{color}
\usepackage{ulem}
\usepackage{physics}
\usepackage{multirow}
\usepackage[table]{xcolor}
\journalname{Earth, Planets and Space}
\begin{document}

\title{
Relationship between three-dimensional velocity of filament eruptions and CME association
}

\titlerunning{3D velocity of filament eruptions \& CME association}        

\author{
Daikichi Seki         \and
Kenichi Otsuji         \and\\
Takako T. Ishii         \and
Ayumi Asai         \and\\
Kiyoshi Ichimoto  
}


\institute{
D. Seki \at
Graduate School of Advanced Integrated Studies in Human Survivability, Kyoto University,
Sakyo, Kyoto 606-8306, Japan	\\
Astronomical Observatory, Kyoto University,
Yamashina, Kyoto 607-8471, Japan	\\
Department of Applied Mathematics and Theoretical Physics, University of Cambridge,
Wilberforce Road, Cambridge, CB3 0WA, United Kingdom		\\
Centre for the Study of Existential Risk, University of Cambridge,
16 Mill Lane, Cambridge CB2 1SB, United Kingdom\\
\email{seki.daikichi.87s@kyoto-u.jp}
\and
K. Otsuji \at
Space Environment Laboratory, Applied Electromagnetic Research Institute, National Institute of Information and Communications Technology,
Koganei, Tokyo 184-8795, Japan
\and
T. T. Ishii,
A. Asai, and
K. Ichimoto       \at
Astronomical Observatory, Kyoto University,
Sakyo, Kyoto 606-8502, Japan
}

\date{Received: date / Accepted: date}
\maketitle

\begin{abstract}
It is widely recognised that filament disappearances or eruptions are frequently associated with Coronal Mass Ejections (CMEs).
Since CMEs are a major source of disturbances of the space environment surrounding the Earth, it is important to investigate these associations in detail for the better prediction of CME occurrence.
However, the proportion of filament disappearances associated with CMEs is under debate.
The estimates range from $\sim$10\% to $\sim$90\% 
and could be affected by the manners to select the events.
In this study, 
we aim to reveal what parameters control the association between filament eruptions and CMEs.
We analysed the relationships between
CME associations and the physical parameters of filaments including their length, maximum ascending velocity, and direction of eruptions using 
28
events of filament eruptions observed in H$\alpha$.
We found that the product of the maximum radial velocity and the filament length is well correlated with the CME occurrence.
If the product is larger than 8.0$\times$10$^{6}$ km$^{2}$ s$^{-1}$, the filament will become a CME with a probability of 93\%, and if the product is smaller than this value, it will not become a CME with a probability of 100\%.
We suggest a kinetic-energy threshold above which filament eruptions are associated with CMEs.
Our findings 
also
suggest the importance of measuring the velocity vector of filament eruption in three-dimensional space for the better prediction of CME occurrence.
\end{abstract}

\section{Introduction}
Filaments are regions of dense cool plasma floating in the corona that are supported by magnetic fields.
They are observed in absorption as dark features on the solar disk in H$\alpha$ (6562.8 \AA) and in emission as prominences above the solar limb.
At the end of its life, a filament disappears by slow fading or exhibits a transient eruption.
Before it disappears or erupts, 
small-scale blobs observed in H$\alpha$ in a filament often show a larger standard deviation of the line-of-sight (LOS) velocity\cite{seki2017increase,seki2019small}.
During eruption phase,
the entire body of a filament ascends at a velocity of 100--1000 km s$^{-1}$ \cite{Parenti_2014}.

Filament eruptions are often associated with coronal mass ejections (CMEs), which are observed by coronagraphs such as the Large Angle and Spectrometric Coronagraph (LASCO) \cite{brueckner1995large}.
Some CMEs exhibit a three-part structure \cite{doi:10.1029/JA090iA01p00275} consisting of a leading edge, faint coronal cavity, and dense core.
Others exhibit more complex forms, appearing as narrow jets or global eruptions, which are called halo CMEs \cite{webb2015eruptive}. 
The core of a CME is believed to originate from the filament mass if the CME is associated with a filament eruption.
Because the Sun is concealed by an occulting disk in coronagraph observations of CMEs, it is difficult to observe the early evolution of CMEs.
Investigations of the evolution of filament eruptions and their association with CMEs 
are expected to
clarify the early evolution of CMEs.

CMEs often produce severe geomagnetic storms, which expose the Earth to a potential risk of adverse socioeconomic impacts such as a widespread blackout\cite{boteler2001assessment}.
A CME associated with a polar crown filament eruption reportedly caused a severe geomagnetic disturbance (Dst $\sim$ $-$200 nT) three days after the eruption\cite{1996JGR...10113497M,mckuro96}.
Therefore, to mitigate the socioeconomic impacts of geomagnetic disturbances, it is essential to predict the occurrence of CMEs and their arrival to the Earth.
And to reveal the relationship between CMEs and the eruption or disappearance of filaments is important for the better prediction.

However, the reported proportion of filament disappearances or eruptions that are associated with CMEs ranges from $\sim$10\% to $\sim$90\%.
%
\cite{hori2002trajectories} studied 50 prominence eruptions observed at 17 GHz by the Nobeyama Radioheliograph \cite{1994IEEEP..82..705N} and found that 92\% of them were associated with CMEs.
\cite{seki20}
investigated 43 filament disappearances in H$\alpha$ data observed by the Solar Dynamics Doppler Imager (SDDI)\cite{ichimoto2017new} on the Solar Magnetic Activity Research Telescope (SMART)\cite{2004ASPC..325..319U} at Hida Observatory, Kyoto University, and found that 50\% of them were associated with CMEs.
\cite{mccauley2015prominence}
studied 904 filament and prominence eruptions observed in He \textrm{II} (304 \AA) by the Atmospheric Imaging Assembly (AIA)\cite{lemen2011atmospheric} and found that 73\% of them were associated with CMEs.
In contrast, 
\cite{Al_Omari_2010}
automatically classified 7332 filament/prominence eruptions reported by the National Centers for Environmental Information\footnote{ftp://ftp.ngdc.noaa.gov/STP/SOLAR\_DATA/SOLAR\_FILAMENTS/ accessed in 2008} as events associated or not associated with CMEs and found that only 17\% of them were associated with CMEs. (For a more detailed summary of previous studies on the filament--CME association, see Table 1 in 
\cite{Al_Omari_2010}). 
It is supposed that the discrepancy among these results could depend on how to select the events.

In this study, we aim to reveal what parameters control the association between the filament eruptions and CMEs.
We investigate the relationships between physical parameters that characterise filament eruptions, i.e., the length, velocity during eruption, and direction of eruption, and the CME association.
Several studies have shown that these parameters are well correlated with the CME association of filament eruptions\cite{seki20,Gilbert_2000,2003PASJ...55.1141M}.
\cite{Gilbert_2000} 
studied 54 prominence eruptions observed above the limb in H$\alpha$.
They defined ``eruptive prominences'' as those in which all or part of the material escaped from the solar gravitational field and ``active prominences'' as those in which none of the material appeared to escape.
They found that eruptive prominences clearly had a larger apparent velocity (the velocity projected on the plane of the sky) above 1.10 solar radii than active prominences did and that eruptive prominences were more strongly associated with CMEs (94\%) than active ones (46\%).
Our previous study\cite{seki20} found that filament eruptions are more likely to be associated with CMEs if the filament length exceeds 150 Mm, the maximum radial velocity exceeds 140 km s$^{-1}$, or their direction is inclined by less than 48 deg with respect to the solar normal.
Thus, in the present study, we focus on these three parameters of filament eruptions and investigate how the association rate varies with respect to them.
Note that, in contrast to our previous work that we investigated the tendency of CME
association with respect to individual physical parameters, the present study aims to improve the predictability of CME association by combining those parameters.
In Section \ref{data}, we provide a description of the data we utilised.
In the succeeding section, the results will be provided, followed by summary and discussion.

\section{Data}\label{data}
We selected events from the SMART/SDDI Filament Disappearance Catalogue\footnote{https://www.kwasan.kyoto-u.ac.jp/observation/event/sddi-catalogue/} (hereafter, the catalogue)\cite{seki20}.
The unique advantage of the SDDI is its wide wavelength coverage, which makes it possible to determine the LOS velocity of erupting filaments up to 400 km s$^{-1}$.
The catalogue lists 43 filament/prominence disappearances observed by SDDI from 2016 May 1 to 2019 June 18, in which filaments/prominences totally disappeared at the H$\alpha$ line centre.
We selected 
28
of these events that had a credibility value of 2 or 3 for CME association in the catalogue (description of ``credibility'' will be provided later).
That is, we used only events whose CME association or non-association is fairly clear.
Note that some of the events were excluded from our analysis, even though their credibility was 2 or 3, because 
(1) terrestrial clouds covered the target filaments, and it was impossible to estimate their precise LOS velocities (No.001, 007, 018, 021, 022, 035, and 043), or 
(2) the length of the target filament could not be measured due to the lack of observation (No.029).
Most of the selected events (26 of 28) are filament eruptions, and two events (on May 24 and June 20 in 2016) are prominence eruptions observed on the solar limb.
Hereafter, we refer to these 28 events simply as filament eruptions.

The credibility value indicates how credible the CME association of an event is.
We made a movie containing solar full-disk images of each event in the H$\alpha$ line centre or in 304\AA\ and LASCO C2 running difference images.
While watching each movie, we examined the directional and temporal association of each filament eruption with the CME and assigned a credibility value based on our judgement.
The actual movies used for examination can be accessed at the catalogue webpage (click the credibility column).
Our judgement is based on (1)$\Delta T$, which is the difference between the time when a CME was first observed in LASCO and the time of total disappearance of the filament in H$\alpha$ centre (same as FD\_end\_time in the catalogue), and (2)$\Delta \phi$, which is
the difference between the position angle of the filament and the central position angle of the CME. 
Table \ref{cmeassociation} describes how we determined the credibility 2 or 3 of the CME association on the basis of $\Delta T$ and $\Delta \phi$.
The credibility 1 was labeled to events which were difficult to determine one-to-one correspondence.
More specifically, we labeled the credibility 1 
(1) if two filaments disappeared within one hour, the difference between their $\Delta \phi$'s was within 15 deg, and they were diagnosed as being associated with the same CME, or 
(2) if there were flares located within 30 deg from the central position angle of the CME and within a few hours prior to CME occurrence.
For example, the events No.002 and No.003 in the catalogue were categorised as the credibility of 1 because two filaments disappeared within one hour, and it was ambiguous which of these events was actually associated with one CME. 
Another example is No.012, in which a CME could be attributable to a C-class flare occurred in an active region rather than to the filament eruption of the interest, and thus we concluded the credibility of this event as 1.
\begin{center}
\begin{threeparttable}
  \caption{
	Criteria for determining the credibility of a CME association on the basis of $\Delta T$ and $\Delta \phi$. The units of $\Delta T$ and $\Delta \phi$ are hour and deg, respectively. 
  }
  \label{cmeassociation}
  \begin{tabular}{c|cccc}
 	&
    $\Delta T < 2$ &
     $2 < \Delta T < 6$&
     $6 < \Delta T < 9$&
    $9 < \Delta T$ \\
    \noalign{\smallskip}\hline\noalign{\smallskip}
	\multirow{2}{*}{$\Delta \phi < 30$} &
		with CME &
		\cellcolor[gray]{0.8}with CME & 
		\cellcolor[gray]{0.4}\textcolor{white}{without CME} &
		\cellcolor[gray]{0}\textcolor{white}{without CME} \\
	&
		credibility 3 & 
		\cellcolor[gray]{0.8}credibility 2 &
		\cellcolor[gray]{0.4}\textcolor{white}{credibility 2} & 
		\cellcolor[gray]{0}\textcolor{white}{credibility 3} \\	
	\multirow{2}{*}{$30 < \Delta \phi < 70$} &
		\cellcolor[gray]{0.8}with CME &
		\cellcolor[gray]{0.8}with CME &
		\cellcolor[gray]{0.4}\textcolor{white}{without CME} &
		\cellcolor[gray]{0}\textcolor{white}{without CME} \\
	&
		\cellcolor[gray]{0.8}credibility 2 &
		\cellcolor[gray]{0.8}credibility 2 &
		\cellcolor[gray]{0.4}\textcolor{white}{credibility 2} &
		\cellcolor[gray]{0}\textcolor{white}{credibility 3}\\
	\multirow{2}{*}{$70 < \Delta \phi < 100$} &
		\cellcolor[gray]{0.4}\textcolor{white}{without CME} &
		\cellcolor[gray]{0.4}\textcolor{white}{without CME} &
		\cellcolor[gray]{0.4}\textcolor{white}{without CME} &
		\cellcolor[gray]{0}\textcolor{white}{without CME} \\
	&
		\cellcolor[gray]{0.4}\textcolor{white}{credibility 2} &
		\cellcolor[gray]{0.4}\textcolor{white}{credibility 2} &
		\cellcolor[gray]{0.4}\textcolor{white}{credibility 2} &
		\cellcolor[gray]{0}\textcolor{white}{credibility 3}\\
	\multirow{2}{*}{$100 < \Delta \phi$} &
		\cellcolor[gray]{0}\textcolor{white}{without CME} &
		\cellcolor[gray]{0}\textcolor{white}{without CME} &
		\cellcolor[gray]{0}\textcolor{white}{without CME} &
		\cellcolor[gray]{0}\textcolor{white}{without CME} \\
	&
		\cellcolor[gray]{0}\textcolor{white}{credibility 3} &
		\cellcolor[gray]{0}\textcolor{white}{credibility 3} &
		\cellcolor[gray]{0}\textcolor{white}{credibility 3} &
		\cellcolor[gray]{0}\textcolor{white}{credibility 3}\\
    \noalign{\smallskip}\hline
  \end{tabular}
  \begin{tablenotes}
    \item[] 
  \end{tablenotes}
\end{threeparttable}
\end{center}

Table \ref{datatable} shows the selected 28 events with their CME associations and physical parameters.
\newpage
\begin{center}
\begin{threeparttable}
  \caption{
	Filament eruptions used in this study.
    Data are taken from the SMART/SDDI Filament Disappearance Catalogue\cite{seki20}.
  }
  \label{datatable}
  \begin{tabular}{ccccccccccc}
    \hline\noalign{\smallskip}
    date \& time&
    CME &
    Cred. \tnote{a}&
    $V_{r\_max}$&
    $V_{r\_fin}$ &
    $V_{pos}$\tnote{b}&
    $V_{max}$ &
    $L$&
    $\Theta$ \tnote{c} \\
    (UT) &
    &
    &
    (km s$^{-1}$)&
    (km s$^{-1}$)&
    (km s$^{-1}$)&
    (km s$^{-1}$)&
    (Mm) &
    (deg) \\
    \noalign{\smallskip}\hline\noalign{\smallskip}
    2016-05-24 01:00 & Yes & 3 & 134  & 36.4 		& 134   & 175  & 137  & 11.8 \\
    2016-06-01 21:00 & Yes & 3 & 173  & 50.8 		& 189   & 226  & 151  & 25.7 \\
    2016-06-20 05:30 & Yes & 3 & 215  & 41.3 		& 223   & 224  & 63.6 & 16.6 \\
    2016-07-07 07:19 & Yes & 3 & 359  & 189 		& 272   & 365  & 35.6 & 9.67 \\
    2016-07-19 05:30 & No  & 3 & 30.8 & $-$3.80 	& 27.0  & 93.1 & 31.7 & 53.9 \\
    2016-07-20 01:00 & Yes & 3 & 122  & $-$6.40 	& 101   & 199  & 99.8 & 43.9 \\
    2016-08-09 23:30 & Yes & 3 & 149  & 51.3 		& 130   & 155  & 449  & 15.0 \\
    2016-08-11 22:00 & No  & 2 & 9.40 & 2.00 		& 38.2  & 52.0 & 153  & 75.9 \\
    2016-08-13 02:30 & No  & 3 & 34.8 & $-$18.0 	& 108   & 234  & 139  & 71.3 \\
    2016-08-19 02:00 & No  & 2 & 3.60 & 2.20 		& 18.1  & 22.3 & 341  & 78.5 \\
    2016-09-03 23:53 & No  & 3 & 15.7 & 15.0 		& 9.35  & 52.7 & 52.5 & 7.89 \\
    2016-09-09 21:51 & No  & 3 & 51.6 & 25.8 		& 22.3  & 154  & 41.7 & 35.2 \\
    2016-11-04 01:32 & Yes & 2 & 38.4 & 26.8 		& 39.6  & 44.3 & 312  & 29.1 \\
    2016-11-05 01:24 & Yes & 3 & 121  & 109 		& 67.0  & 123  & 112  & 7.26 \\
    2017-02-10 04:00 & No  & 2 & 4.90 & $-$1.40 	& 11.5  & 14.2 & 154  & 64.9 \\
    2017-02-19 04:44 & Yes & 2 & 161  & 143 		& 185   & 218  & 112  & 42.2 \\
    2017-03-05 01:30 & No  & 3 & 68.1 & 40.9 		& 74.4  & 81.0 & 42.7 & 32.7 \\
    2017-04-23 04:30 & Yes & 3 & 456  & 436 		& 497   & 523  & 113  & 29.4 \\
    2017-04-23 03:05 & No  & 3 & 18.9 & $-$25.0 	& 18.1  & 324  & 73.9 & 45.5 \\
    2017-04-23 23:08 & Yes & 3 & 80.3 & 36.5 		& 110   & 112  & 293  & 44.4 \\
    2017-04-29 23:30 & Yes & 2 & 44.2 & $-$10.6 	& 151   & 154  & 271  & 73.0 \\
    2017-09-25 21:24 & Yes & 2 & 172  & 7.70 		& 530   & 555  & 164  & 72.0 \\
    2017-12-07 02:15 & No  & 3 & 2.30 & $-$43.3 	& 63.8  & 246  & 69.8 & 88.0 \\
    2018-04-03 02:30 & No  & 3 & 82.7 & $-$2.77 	& 139   & 183  & 160  & 59.0 \\
    2018-04-20 05:00 & No  & 2 & 0.100& $-$23.4 	& 0.141 & 46.7 & 59.3 & 72.5 \\
    2018-07-16 06:00 & No  & 3 & 40.0 & $-$5.00 	& 52.8  & 75.6 & 77.4 & 43.9 \\
    2018-07-31 03:00 & No  & 3 & 3.70 & 3.30 		& 0.412 & 13.5 & 99.6 & 26.1 \\
    2019-02-23 22:30 & Yes & 3 & 47.6 & 28.2 		& 45.1   & 66.6 & 285  & 43.6 \\
    \noalign{\smallskip}\hline
  \end{tabular}
  \begin{tablenotes}
    \item[a] The credibility of the association between a CME and a filament eruption.
      3 $>$ 2 $>$ 1.
      Events with a credibility of 1 are excluded from this study.
    \item[b] The apparent velocity of a filament.
      It is defined as $\sqrt{V_{x}^2 + V_{y}^2}$, where $V_{x}$ and $V_{y}$ are the east--west and south--north velocities projected on the plane of the sky, respectively.
    \item[c] The inclination angle of a filament eruption with respect to the solar normal.
    \item[]
  \end{tablenotes}
\end{threeparttable}
\end{center}
`Date \& time' is the start time of a filament eruption and is defined as the first observation of a dark feature in H$\alpha$ $-$ 0.5 \AA.
It is equivalent to `FD\_start\_time' in the catalogue.
`CME' indicates whether a filament eruption is associated with a CME.
$V_{r\_max}$, $V_{r\_fin}$, $V_{pos}$,
and $V_{max}$
 are determined as follows.
We manually tracked and measured the position and LOS velocity of a blob at the apex of the filament that was present until its total disappearance in H$\alpha$.
Then, we constructed its three-dimensional velocity as a function of time.
$V_{r\_max}$ is the maximum radial (or ascending) velocity during the eruption,
whilst $V_{max}$ is the maximum magnitude of three-dimensional velocity.
$V_{pos}$ is equal to $\sqrt{V_{x}^2 + V_{x}^2}$, where $V_{x}$ and $V_{y}$ are the velocities of the filament in the east--west and south--north directions on the plane of the sky, at the time of $V_{r\_max}$, respectively.
$V_{r\_fin}$ is the radial velocity at the last observation of a filament in H$\alpha$.
$L$ is the length of a filament measured at the same time as `date \& time'.
The projection effect is corrected according to the location of the filament on the solar disk.
$\Theta$ is the inclination angle between the direction of the filament velocity at the time of $V_{r\_max}$ and the solar normal (see Figure B on the catalogue webpage).
For further details of how these values were determined, see our previous paper\cite{seki20}.

\section{Results}
\begin{figure}
  \includegraphics[width=210bp, bb=10 10 500 400]{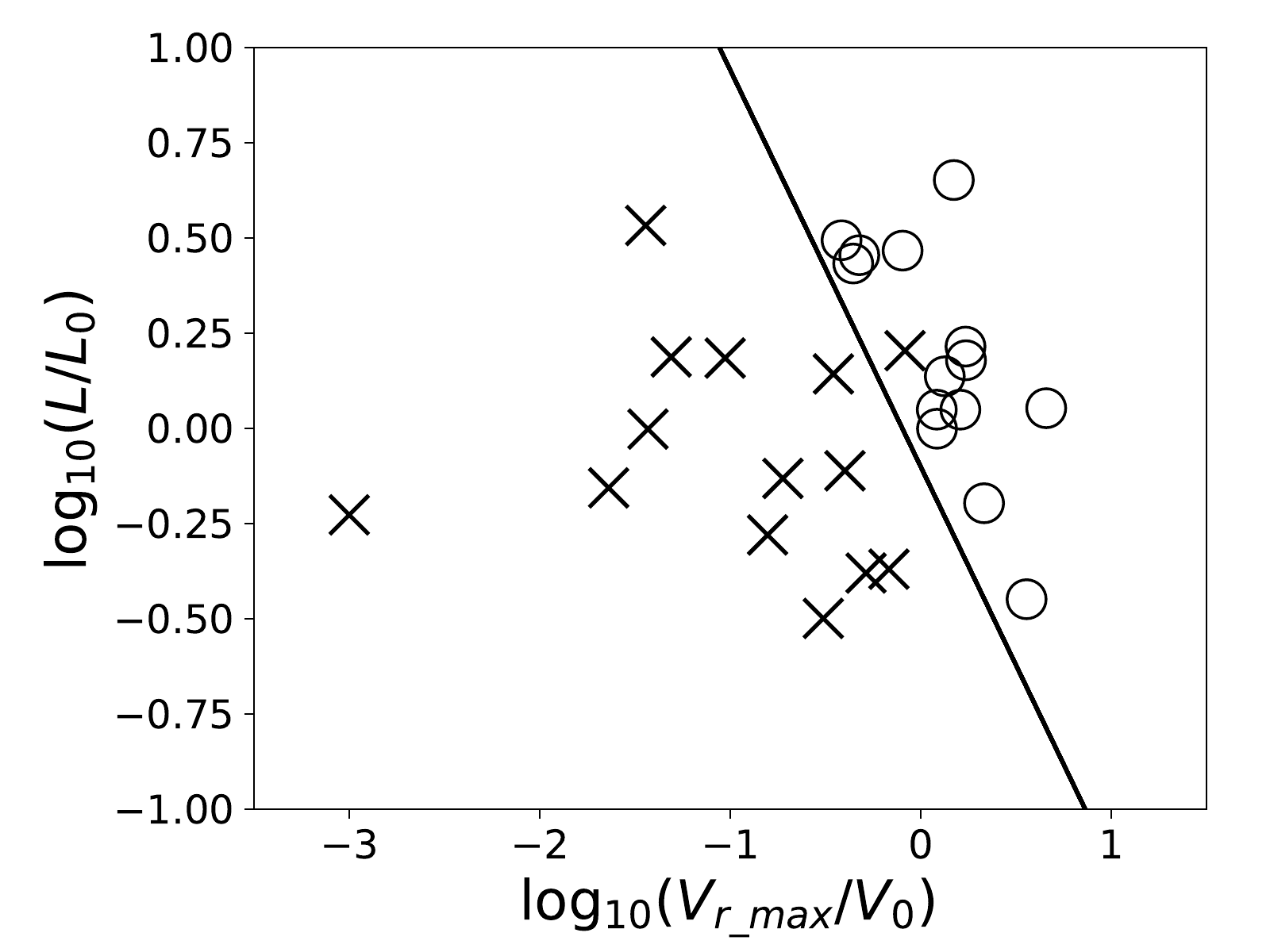}
  \includegraphics[width=210bp, bb=10 10 500 400]{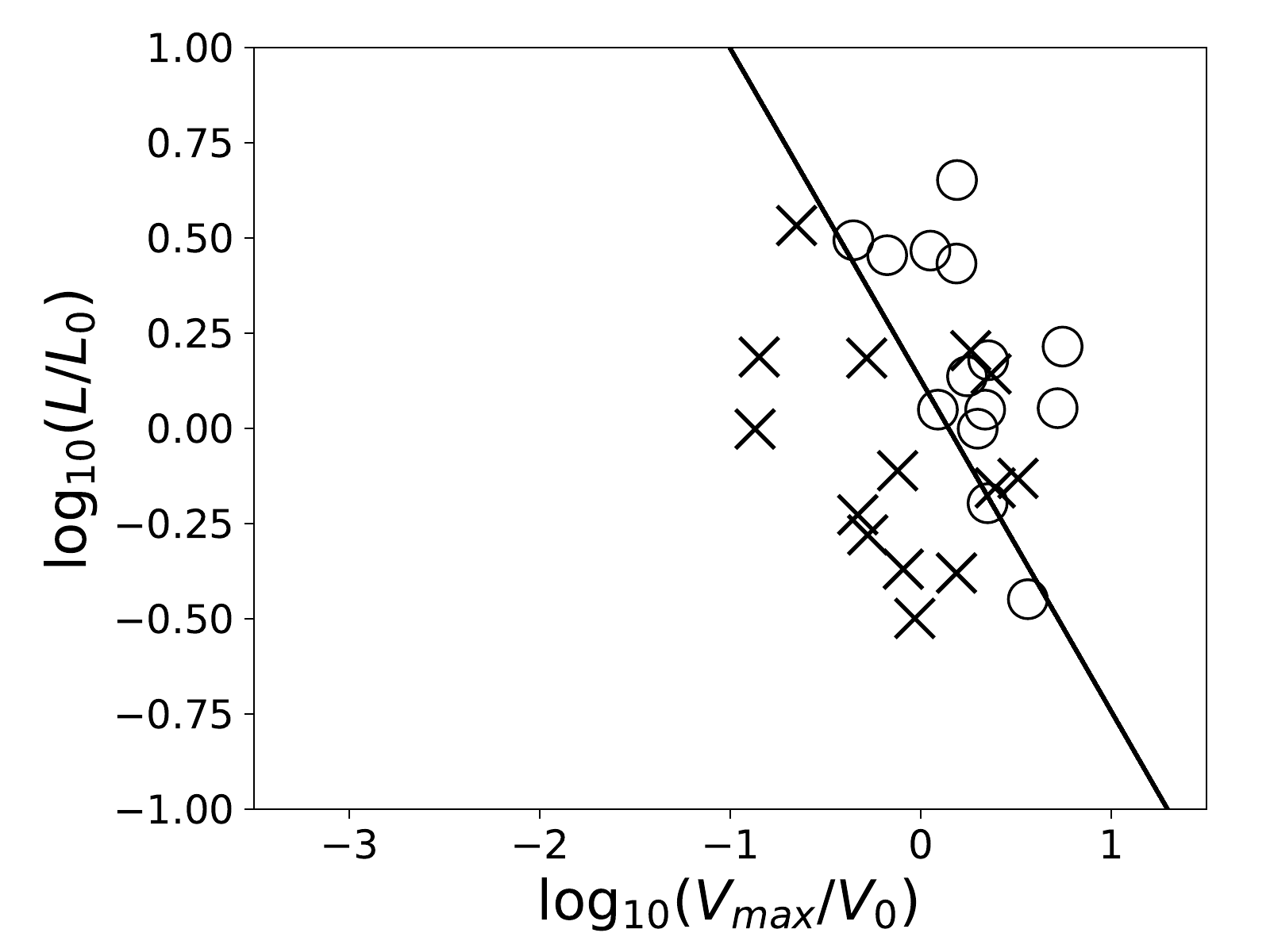}
   \includegraphics[width=210bp, bb=10 10 500 400]{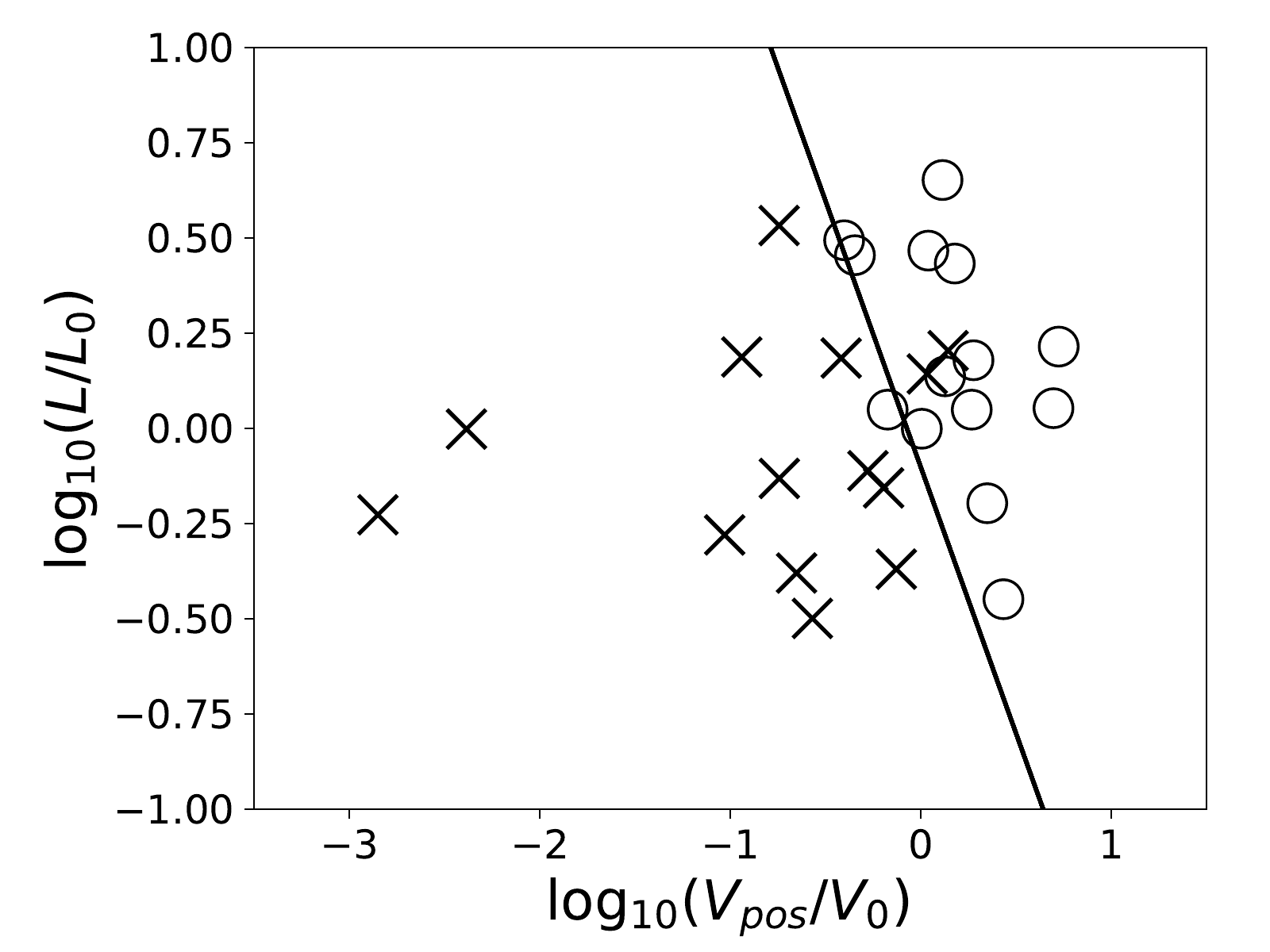}
  \caption{
  Plots of filament eruptions according to $V_{r\_max}$ (top left), $V_{max}$ (top right), or $V_{pos}$ (bottom left) and filament length, $L$, on a common logarithmic scale.
  $V_{0}$ and $L_{0}$ correspond to the typical velocity (100 km s$^{-1}$) and typical length (100 Mm) of filaments, respectively.
  Open circles and crosses represent events with and without CMEs, respectively.
  The solid lines are described in the text.
  }
  \label{vrmax-len}
\end{figure}

Figure \ref{vrmax-len} displays the CME associations according to $L$ (vertical axis) and $V_{r\_max}$ (top left), $V_{max}$ (top right), or $V_{pos}$ (bottom left) on a logarithmic scale.
Here, the length and velocities are normalised by $L_{0} = 100$ Mm and $V_{0} =$ 100 km s$^{-1}$, respectively.
We can see the tendency that the longer and faster filaments are more likely to be associated with CMEs.
The solid lines in the panels are drawn by the following relationships;
\begin{eqnarray}
\left(\frac{V_{r\_max}}{V_{0}}\right) \times \left(\frac{L}{L_{0}}\right)^{0.96} = 0.80 \label{vrmax}\\
\left(\frac{V_{max}}{V_{0}}\right) \times \left(\frac{L}{L_{0}}\right)^{1.1} = 1.4 \label{vmax}\\
\left(\frac{V_{pos}}{V_{0}}\right) \times \left(\frac{L}{L_{0}}\right)^{0.72} = 0.85 \label{vpos}
\end{eqnarray}
They were determined by using the algorithm of Linear Support Vector Classification implemented in LIBLINEAR\cite{fan2008liblinear} (for further explanation of the algorithm, see Appendix and 
\cite{fan2008liblinear}).
In the top left panel ($V_{r\_max}$), 
27 events out of 28 (96\%)
were correctly classified into the two groups of filament eruptions with (open circles) and without (crosses) CMEs, whilst in the other cases, 
21 (for $V_{max}$) and 25 ($V_{pos}$)
events were correctly separated.
If we make a separation so that the number of correctly classified events can be maximised, 27 ($V_{r\_max}$), 24 ($V_{max}$), and 26 ($V_{pos}$) events will be correctly classified (the separations not shown in the figure).
Thus, a better prediction of the CME association could be obtained by using $V_{r\_max}$ rather than using $V_{max}$ or $V_{pos}$ at least with our limited number of the events, 28.
This result suggests the advantage of measuring the radial velocity of filament eruptions.
It also suggests that measuring both the velocity and the length of filaments should contribute to the better prediction of CME occurrence.
The three-dimensional velocity observation provides a better capability for predicting the occurrence of CMEs, whilst the H$\alpha$ imaging observations without Doppler measurements ($V_{pos}$) can still contribute to it.

\begin{figure}
  \includegraphics[width=210bp, bb=10 10 500 400]{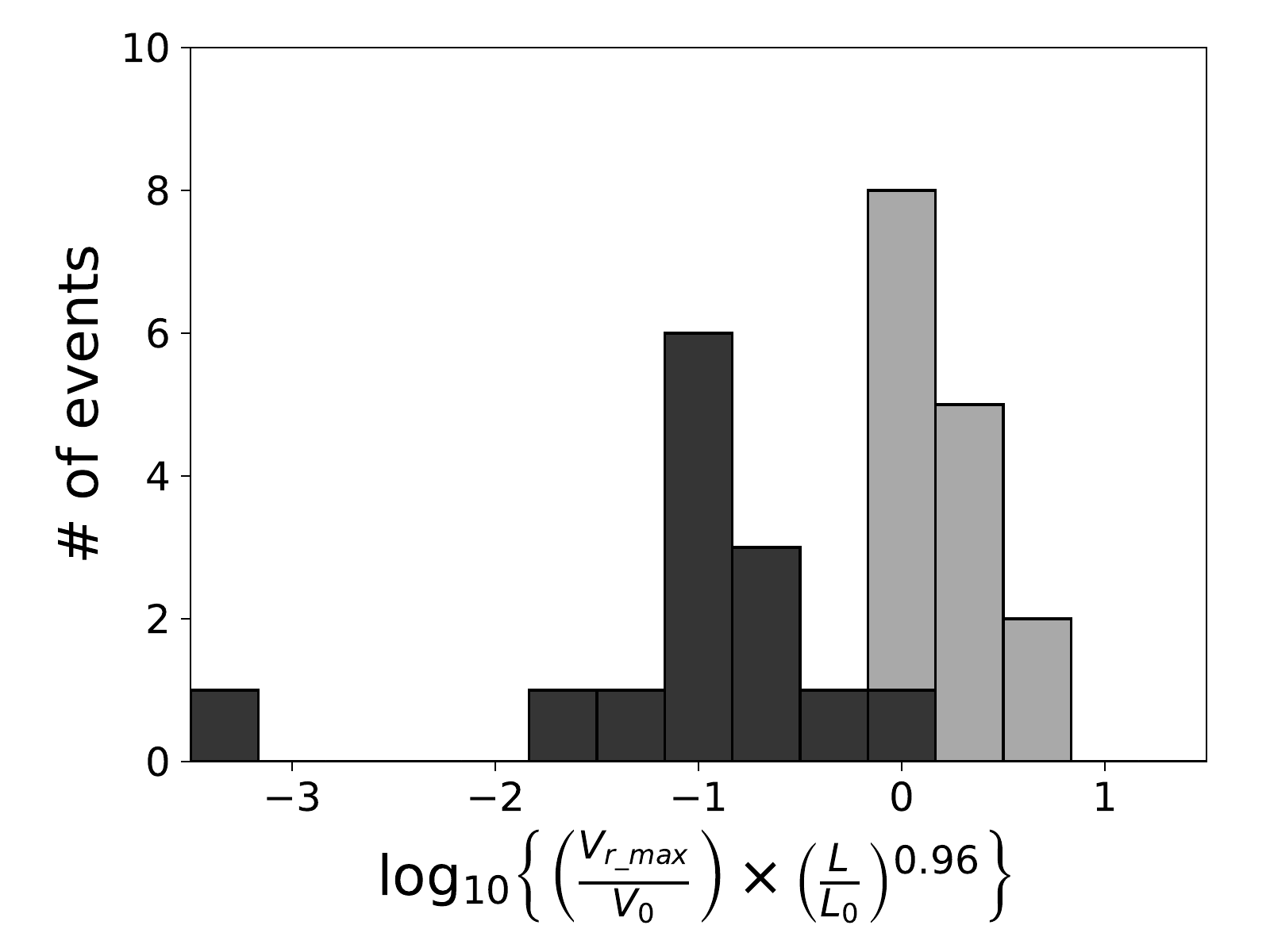}
  \includegraphics[width=210bp, bb=10 10 500 400]{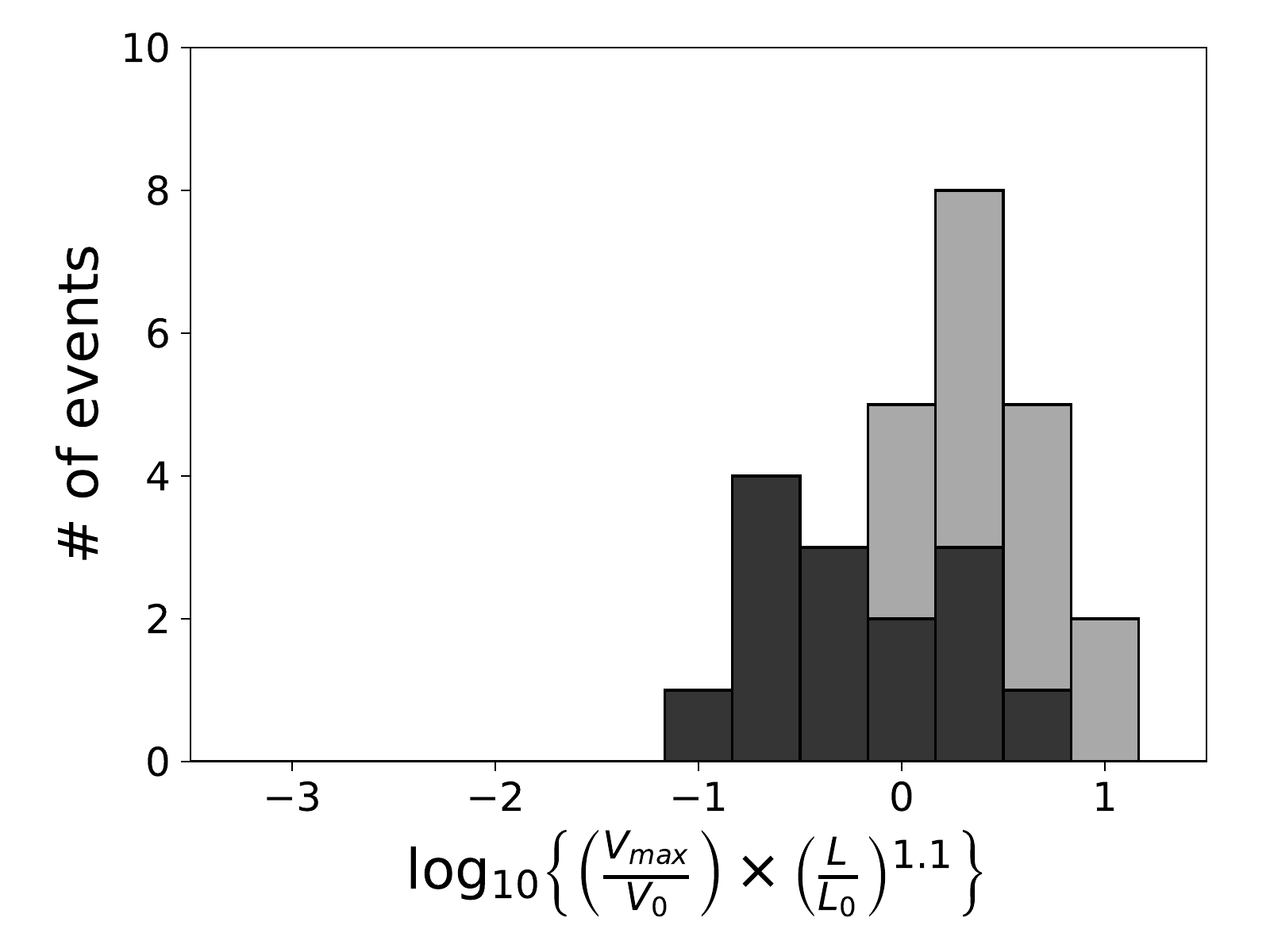}
  \includegraphics[width=210bp, bb=10 10 500 400]{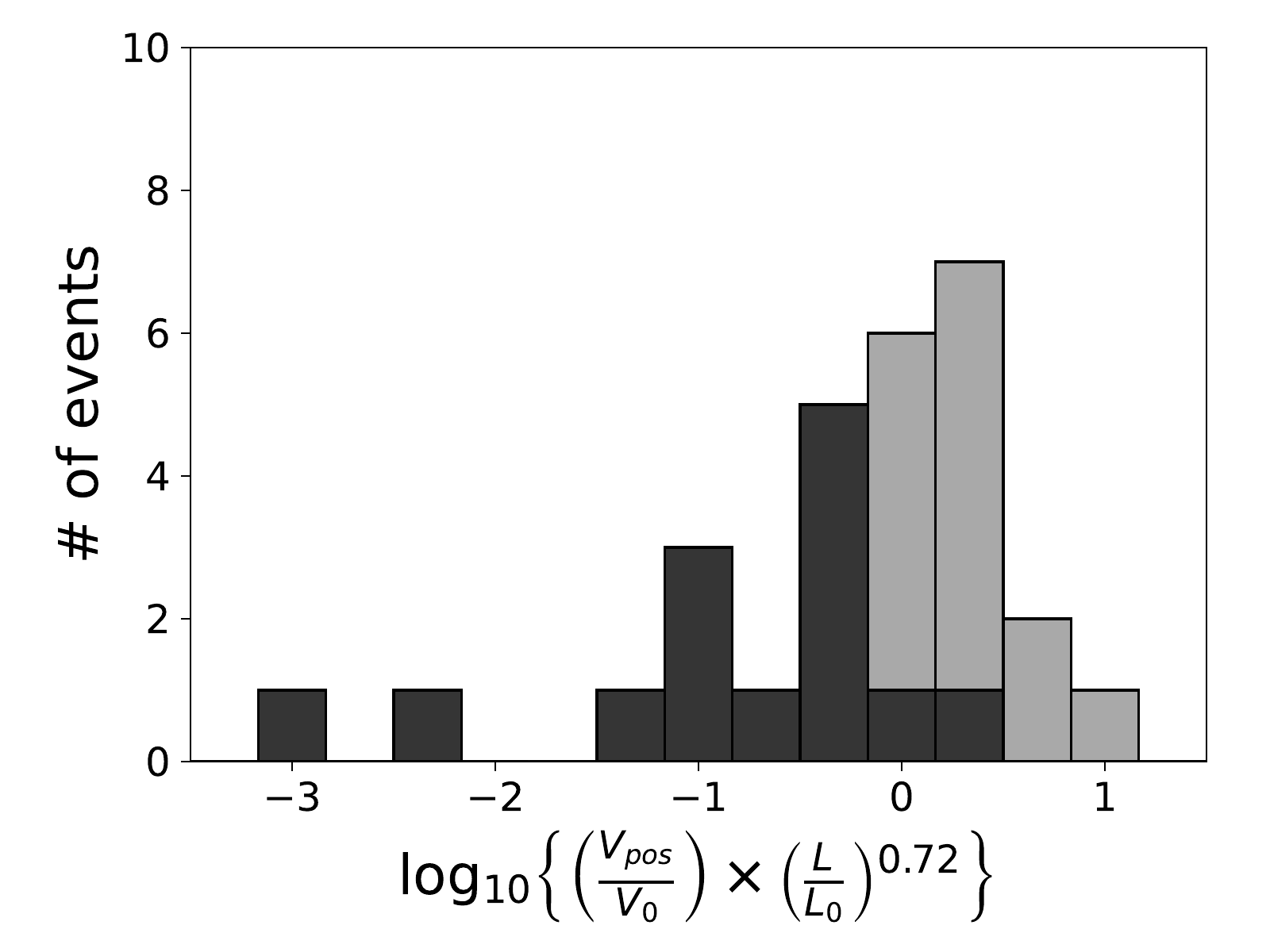}
  \caption{
  Histograms of the LHS of Equation (\ref{vrmax})--(\ref{vpos}) on a common logarithmic scale.
  The dark and light grey bars correspond to the events without and with CMEs, respectively.
  The bars are stacked.
  The size of a bin is 0.33
  }
  \label{vrlen_hist}
\end{figure}

Figure \ref{vrlen_hist} displays histograms of events with (grey) and without (black) CMEs with respect to the left-hand sides (LHS) of Equation (\ref{vrmax})--(\ref{vpos}).
These histograms also demonstrate that the CME association is better identified when $V_{r\_max}$ is used than when $V_{max}$ or $V_{pos}$ is used.
In the top left panel, we see a clear bimodal distribution, which is less clear in the other cases.
To confirm the bimodality quantitatively, we introduced a statistic, D-value, which is defined as 
\begin{equation}
  D \equiv \frac{|\mu_1 - \mu_2|}{\sigma},
\end{equation}
where $\mu_1$ and $\mu_2$ are the averages of the two normal distributions fitted to the events with and without CMEs, and $\sigma$ is equivalent to
\begin{equation}
  \sigma = \sqrt{\frac{\sigma_1^2 + \sigma_2^2}{2}},
\end{equation}
where $\sigma_1$ and $\sigma_2$ are their standard deviations.
The D-value represents the distance between the means of two normal distributions relative to their standard deviations.
These distributions can be regarded as being separated if the D-value is larger than 2 \cite{1994AJ....108.2348A,Muratov_2010,Carlson_2018}.
The means and standard deviations of two normal distributions (with and without CMEs) for each case of Equation (\ref{vrmax})--(\ref{vpos}) are summarised in Table \ref{dvalue} together with the D-values.
We obtained $D$ values 
of 
2.3, 1.8, and 1.7
for $V_{r\_max}$, $V_{max}$, and $V_{pos}$, respectively.
Only $V_{r\_max}$ exhibits the D-value larger than 2.
The better bimodality when $V_{r\_max}$ is used is confirmed quantitatively.
\begin{center}
\begin{threeparttable}
  \caption{
	Summary statistics of the fitted normal distributions and the corresponding $D$ values.
  }
  \label{dvalue}
  \begin{tabular}{cccccc}
    \hline\noalign{\smallskip}
    & \multicolumn{2}{c}{with CMEs} & \multicolumn{2}{c}{without CMEs} & \\
    &  $\mu_{1}$ & $\sigma_{1}$ & $\mu_{2}$ & $\sigma_{2}$ & $D$\\
    \noalign{\smallskip}\hline\noalign{\smallskip}
      $V_{r\_max}$ (Eq. (\ref{vrmax}))& 0.28 & 0.23 & -1.0 & 0.76 & 2.3 \\
    $V_{max}$ (Eq. (\ref{vmax}))& 0.46 & 0.28 & -0.20 & 0.44 & 1.8 \\
  $V_{pos}$ (Eq. (\ref{vpos})) & 0.29 & 0.29 & -0.81 & 0.88 &1.7 \\
    \noalign{\smallskip}\hline
  \end{tabular}
  \begin{tablenotes}
    \item[]
  \end{tablenotes}
\end{threeparttable}
\end{center}

\begin{figure}
   \includegraphics[width=210bp, bb=10 10 500 400]{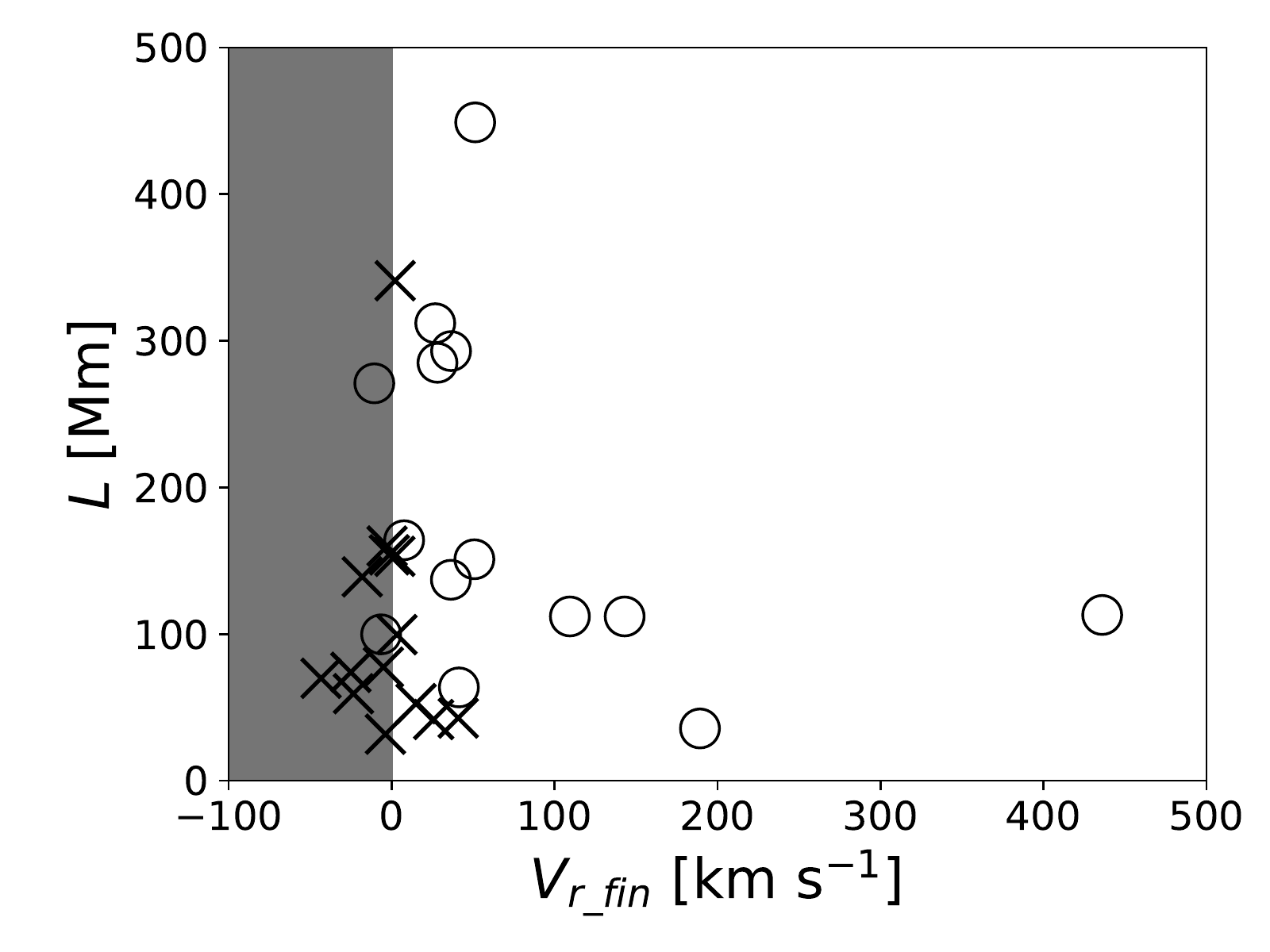}
      \includegraphics[width=210bp, bb=10 10 500 400]{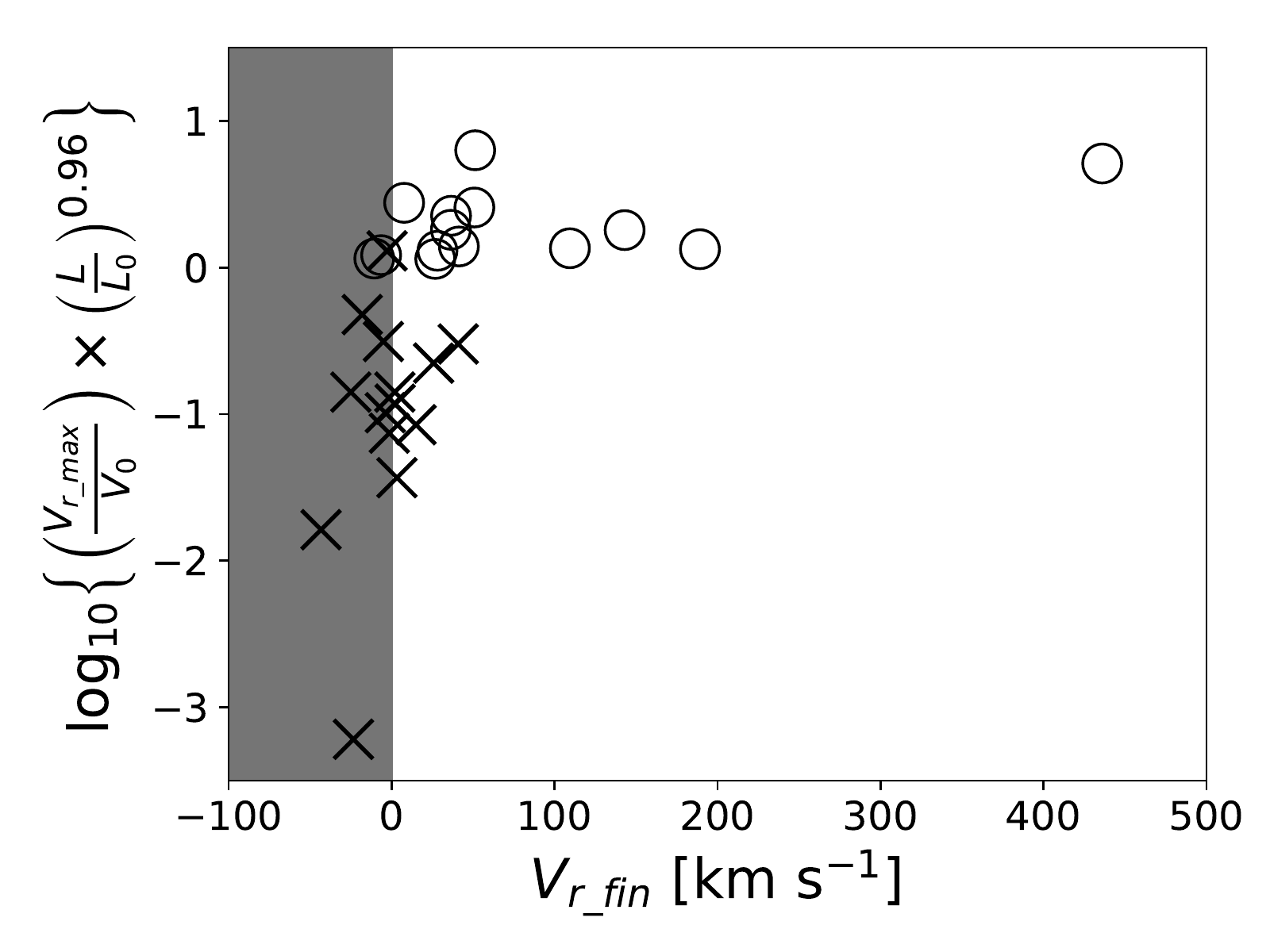}
  \caption{
  Plot of filament eruptions according to $V_{r\_fin}$ and $L$ (left) or common log of the product of normalised $V_{r\_max}$ and normalised $L$ to the power of 0.92 (right).
  Open circles and crosses have the same meaning as in Figure \ref{vrmax-len}.
  Grey area corresponds to negative $V_{r\_fin}$.
  }
  \label{vrfin-len}
\end{figure}

Figure \ref{vrfin-len} shows the CME association with respect to the radial velocity of the last observation ($V_{r\_fin}$) and $L$ (left panel) or a common log of the LHS of Equation (\ref{vrmax}) (right panel) for each filament eruption.
Open circles and crosses denote events with and without CMEs, respectively.
Most of the filament eruptions 
(80\%)
with negative $V_{r\_fin}$ (grey area), i.e., events in which the filaments fall back to the Sun, were not associated with CMEs.
In addition, 
77\%
of the filament eruptions with positive $V_{r\_fin}$ and $L$ larger than 70 Mm were associated with CMEs.
Note that the filaments with the smaller (larger) value of the LHS of Equation (\ref{vrmax}) similarly tend to have smaller (larger) $V_{r\_fin}$ (see the right panel).

From Figure \ref{vrfin-len}, we can also recognise that there are exceptional events that were associated with CMEs despite their negative $V_{r\_fin}$'s ($-$6.40 km s$^{-1}$ and $-$10.6 km s$^{-1}$).
We speculate that the blobs which escaped the solar gravity and erupted into the interplanetary space became invisible in H$\alpha$, and we tracked a part of the filament that fell back to the solar surface.

\begin{figure}
  \includegraphics[width=400bp, bb=10 10 500 350]{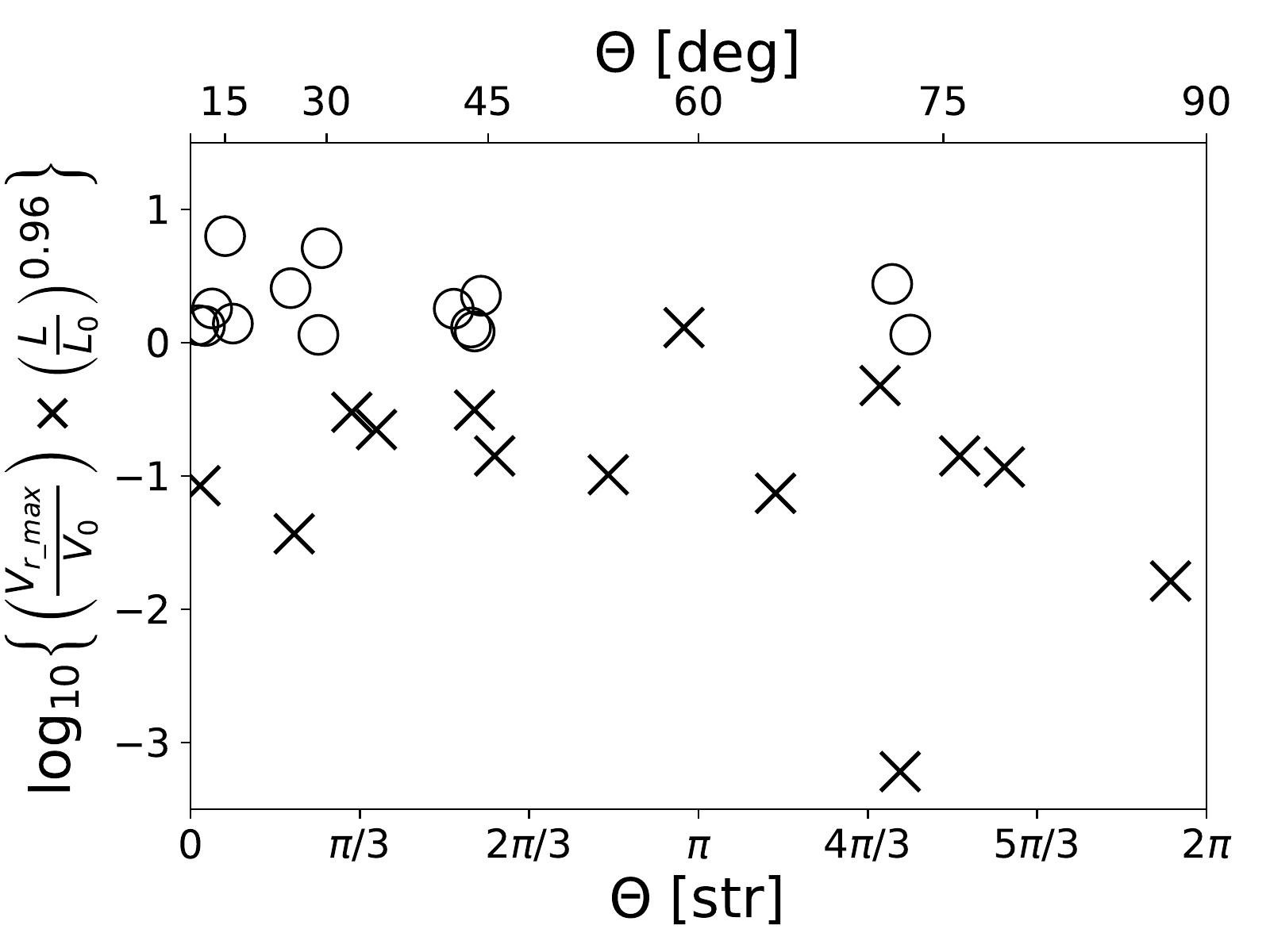}
  \caption{
    Plot of filament eruptions according to $\Theta$ (inclination angle in steradians and degrees) and a common log of the LHS of Equation (\ref{vrmax}).
    The symbols have the same meaning as in Figure \ref{vrmax-len}.
  }
  \label{solang-vrlen}
\end{figure}

Figure \ref{solang-vrlen} shows the CME association according to the LHS of Equation (\ref{vrmax}) (vertical axis) and $\Theta$, the inclination angle (angle from the solar normal) of the velocities (horizontal axis).
We found that 
82\%
of the filament eruptions with directions that were inclined by more than 45 deg from the solar normal were not associated with CMEs, and 
71\%
of those with their $\Theta$'s smaller than 45 deg were associated with CMEs.
Thus, the inclination angle of eruptions will provide a clue for forecasting the CME occurrence. 
Note that the LHS of Equation (\ref{vrmax}) of the filament eruptions with their $\Theta$'s larger than 45 deg seldom exceeds $-$0.097 (= log10(0.80)).
Figure \ref{solang-vrlen} also shows that 
86\%
of the events associated with CMEs have their $\Theta$'s smaller than 45 deg,
while only 36\% of the non-associated ones have $\Theta$ smaller than 45 deg.
These results are consistent with the work of \cite{gopalswamy2003prominence}, in which they defined two types of prominence eruptions, radial and transverse events, according to the eruptive motion of the prominence observed on the solar limb.
In their statistical study, they found that 94\% of the prominence eruptions associated with CMEs were radial events (86\% in our study) and that 76\% of the transverse events were not associated with CMEs (82\% in our study).

\begin{figure}
  \includegraphics[width=8cm, bb=9 9 258 344]{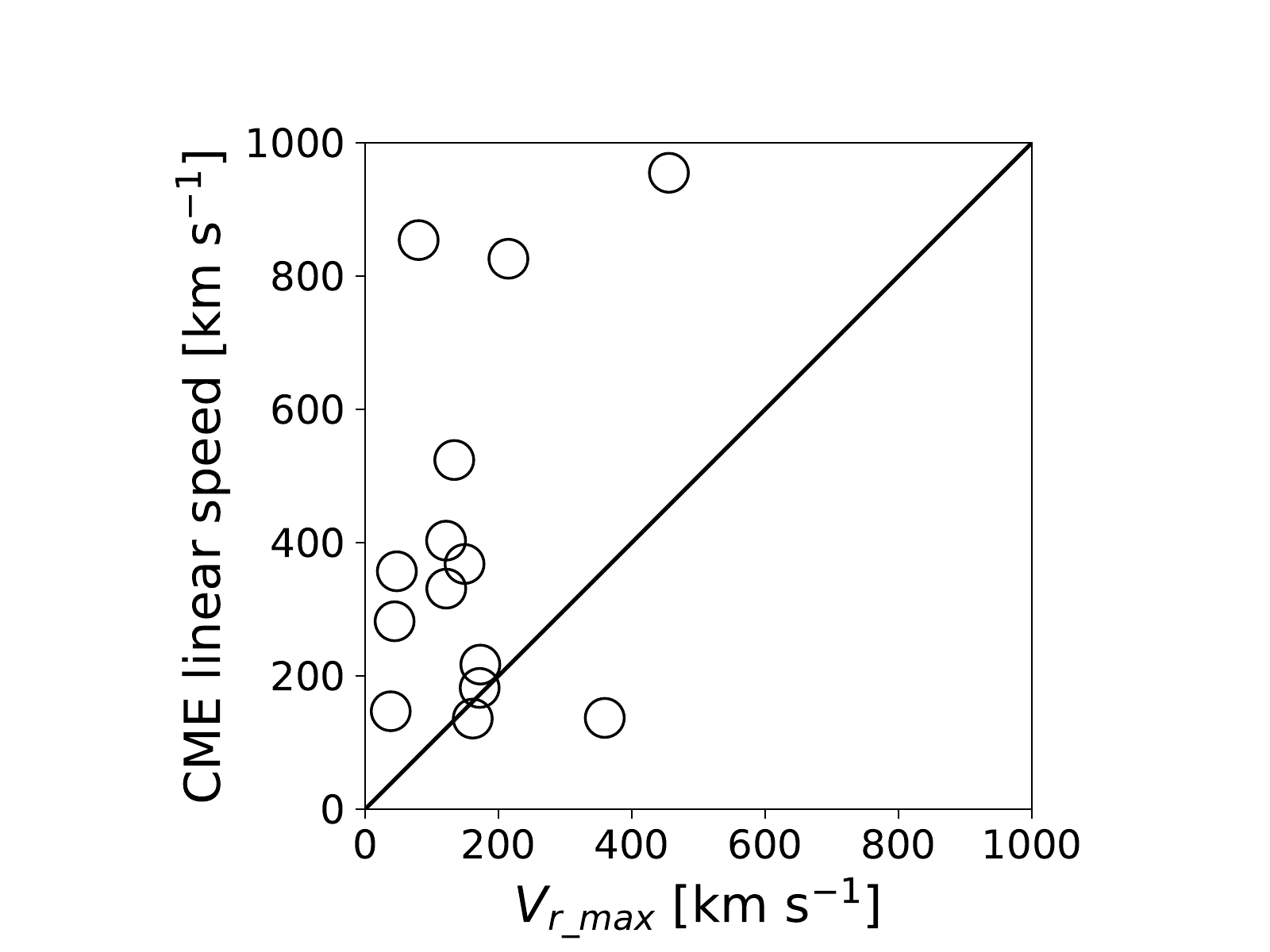}
  \caption{
    Plot of filament eruptions according to $V_{r\_max}$ 
    and CME linear speed.
    The black line corresponds to equal speeds.
  }
  \label{cmespeed-vrmax}
\end{figure}

Figure \ref{cmespeed-vrmax} shows the linear speed of the CMEs in the SOHO/LASCO CME Catalog\cite{2004JGRA..109.7105Y} against $V_{r\_max}$.
The CME speed is expected to be larger than the velocity of the associated filament eruption\cite{gopalswamy2003prominence}.
The black line indicates equal speeds.
All the data points but one ($V_{r\_max}$ $=$ 359 km s$^{-1}$) are above or close to the black line, as expected.
However, since there is a large scatter in the ratio of CME linear speed to $V_{r\_max}$, it is difficult to predict CME linear speeds from the maximum radial velocity.

\section{Summary and Discussion}
In this study, we investigated the relationships between the physical parameters of filament eruptions (three-dimensional velocity, filament length, and direction of eruption) and their CME associations using
28
events observed by SDDI at Hida Observatory.
We found that the filament eruptions are 
well separated into two groups of events, one with and the other without CMEs, according to the product of the normalised maximum ascending velocity ($V_{r\_max} / V_{0}$) and the normalised filament length ($L / L_{0}$) to the power of 0.96, and that among the filament eruptions with $\left(\frac{V_{r\_max}}{V_{0}}\right) \times \left(\frac{L}{L_{0}}\right)^{0.96} > 0.80$, 93\% are associated with CMEs, and 100\% of filament eruptions with the product $< 0.80$ are not associated with CMEs.
The apparent velocity and the length of filaments measured in H$\alpha$ observation could also provide a good measure for predicting the occurrence of CMEs, though the accuracy of the prediction using the apparent velocity is worse than that using the radial velocity.
Our results suggest that the three-dimensional velocity, or more specifically the radial velocity derived from it, and the length of the erupting filament are the notable parameters for improving the predictability of CME association.
And thus, we suggests the importance of observations of the three-dimensional velocity of filament eruptions for the prediction of CMEs.
It should be noted, however, that improvement of statistics, i.e., studies with a larger number of examples, are strongly required to confirm these results.

Here, we propose a possible physical interpretation for the solid line in the top left panel of Figure \ref{vrmax-len}.
This line, which is represented by Equation (\ref{vrmax}), successfully separates events into those with and without CMEs.
We assume that 
(1) the cross section of filaments, $A$, follows the relationship of
\begin{equation}
\left(\frac{A}{A_{0}}\right) = \left(\frac{L}{L_{0}}\right),
\end{equation}
where $A_{0}$ is the typical cross section of filaments (100 Mm$^{2}$), 
and that (2) the average hydrogen density is common among filaments, i.e., 10$^{11}$ cm$^{-3}$, which is a typical value for quiescent prominences \cite{Heinzel_2008}.
Then, if we regard Equation (\ref{vrmax}) as
\begin{equation}
\left(\frac{V_{r\_max}}{V_{0}}\right) \times \left(\frac{L}{L_{0}}\right) \sim 0.80,
\end{equation}
or $V_{r\_max} \times L \sim 8.0 \times 10^{6}\ \mathrm{km}^{2}\ \mathrm{s}^{-1}$, then
its square represents the kinetic energy of an erupting filament, i.e., $\frac{1}{2}\times$ proton mass $\times$ density $\times$ volume $\times$ Vr\_max$^{2}$ =  
5.4 $\times$ 10$^{28}$ erg.
This relationship could be regarded as the kinetic-energy threshold above which filament eruptions are associated with CMEs.
Note that if the length of a filament is 100 Mm, the deduced mass gets 1.7 $\times$ 10$^{15}$ g.
\cite{2006ApJ...641..606G} 
reported the masses of 18 prominences, which ranged from (1.08 $\pm$ 0.52)$\times$10$^{14}$ to (2.09 $\pm$ 0.80)$\times$10$^{15}$ g.
Our assumed ``typical'' mass is consistent with the reported values.

As mentioned in Section 1, the CME association rates of filament eruptions reported to date range from $\sim$10\% to $\sim$90\%.
Here, we provide a possible interpretation of this wide range based on our results.
We showed that the product of the normalised radial velocity of eruptions and the normalised filament length makes a key contribution to the CME association.
The high association rates of 80\%--90\% in the past studies might be attributable to the criteria they used to select the events, under which the prominences have a predominantly large radial velocity and a large size.
\cite{Gilbert_2000} 
reported that 94\% of eruptive prominences (for the definition, see Section 1) were associated with CMEs.
\cite{gopalswamy2003prominence} 
also reported that 83\% of radial prominence eruptions were associated with CMEs.
Their selected prominence eruptions should have had a predominant radial velocity.
In addition, 
\cite{gopalswamy2003prominence} and \cite{hori2002trajectories} 
detected prominences with the NoRH that has its spatial resolution of 10 arcsec\cite{1994IEEEP..82..705N}, which is worse than the spatial sampling of the SDDI (1.23 arcsec pixel$^{-1}$).
Therefore, the selected prominences in these studies seem to have a larger size (e.g., larger than 70 Mm, because 
75\%
of the filaments smaller than 70 Mm were not associated with CMEs according to our result.)

The association rate could also depend on whether studies include disk events (filament disappearances) in the records.
In contrast to the high association rates (80 to 90\%) reported in the studies taking into account only limb events (prominence disappearances)\cite{Gilbert_2000,gopalswamy2003prominence,hori2002trajectories}, 
some studies\cite{Pojoga_2003,2004ApJ...614.1054J,seki20} in which both disk and limb events were considered manifested the association rate of approximately 40--50\%.
\cite{Pojoga_2003} 
reported that 39\% of filament and prominence eruptions observed in H$\alpha$ were associated with CMEs.
\cite{2004ApJ...614.1054J}
reported that 56\% of filament eruptions were associated with CMEs by  automatically detecting filament disappearances in H$\alpha$.
In our study, considering only credible events, we found that 
50\%
of filament eruptions in H$\alpha$ were associated with CMEs.

Additionally, the observational wavelengths at which filaments or prominences are detected could also affect the association rate.
In H$\alpha$, as mentioned in the previous paragraph, approximately 40 to 50\% of disappearance events were associated with CMEs.
By contrast, 
\cite{mccauley2015prominence}
used full-disk solar images in the 171, 193, and 304 \AA\ AIA passbands and reported an association rate of 72\%.

The low association rate (17\%, 
\cite{Al_Omari_2010}
) might be attributable to the fact that the authors include ejecta such as surges in addition to filament eruptions in their sample.
Among their 7332 events, they introduced 15 ``filament types'', including coronal rain, sprays, and surges.
In our study, we did not refer to these ejecta as filaments, and we excluded them from our list.
Thus, the definition of filaments in that study was different from ours.
Moreover, most of their events ($\sim$80\%) were smaller than $\sim$70 Mm (see Figure 8 in 
\cite{Al_Omari_2010}).
According to our result (see Figure \ref{vrmax-len} or \ref{vrfin-len}), 
75\%
of the eruptions of filaments with lengths smaller than 70 Mm were not associated with CMEs.
Assuming that this relation holds for coronal rain, surges, and sprays, 
$\sim$60\% (80\% $\times$ 75\%)
of all their selected events may not be associated with CMEs in our criteria.
Therefore, the low association rate in their study can also be attributed to the event selection criteria; i.e., a significant portion of their events is thought to be located below the threshold line proposed in this paper.

The results of this study can be used to develop a methodology to predict the occurrence of CMEs by measuring the three-dimensional velocities of filament eruptions.
Moreover, our previous works suggest that the occurrence of filament eruptions can be predicted prior to their initiation by 1.3 $\pm$ 0.47 hour for intermediate filaments, on the basis of the mean and standard deviation of the LOS velocity distribution in filaments \cite{seki2017increase,seki2019small}.
Hence, by using SDDI data and measuring the LOS velocity of filaments, we could predict the occurrence of filament eruptions $\sim$1 hour in advance and also, during eruptions, estimate the possibility of CME association before coronagraph observations.

\section*{Appendix Linear Support Vector Classification (LSVC)}
Linear Support Vector Classification (LSVC), which we utilised to estimate the CME association from the observation of filaments, is one of the popular machine-learning methods for classification.
Our goal is to obtain the coefficients of the solid lines in Figure \ref{vrmax-len}, which successfully separate the events associated with CMEs from those without CMEs in accordance with the velocity and length of filaments.
These lines can be expressed as
\begin{eqnarray}
\vb*{w}^{T}\vb*{x} = w_{0} + w_{1}x_{1} + w_{2}x_{2} = 0,
\end{eqnarray}
where $\vb*{w} = (w_{0}, w_{1}, w_{2})^{T}$ is a coefficient vector to be optimised, and 
$\vb*{x} = (1, x_{1}, x_{2})^{T}$ is a feature vector, which corresponds to the observation.
In our case, $x_{1}$ is a common log of a normalised velocity ($V_{r\_max}$, $V_{max}$, or $V_{pos}$ divided by $V_{0}$), and $x_{2}$ is a common log of a normalised length ($L$ divided by $L_{0}$).

We optimised $\vb*{w}$ by minimising the loss, $l$, defined as
\begin{eqnarray}
l = \frac{1}{2}\vb*{w}^{T}\vb*{w} + C\sum^{N}_{i=1} (\max(1 - y^{i}\vb*{w}^{T}\vb*{x}^{i}, 0))^{2},
\label{loss}
\end{eqnarray}
where $i$ and $N$ are the index and the number of our selected events, $y^{i} \in \{-1, 1\}$ is the label of the CME association for the event $i$ (-1: without CME, 1: with CME), $\vb*{x}^{i} = (1, x_{1}^{i}, x_{2}^{i})^{T}$ is an actual observed values for the event $i$, and $C$ is a constant (in our case, set to be 100).
The first term in Equation (\ref{loss}) is a penalty term, which prevents the classifier from overfitting to the sample.
As for the second term, intuitively, to minimise it corresponds to (1) maximising the sum of the distances to the line from the correctly classified ``near'' data points and, simultaneously, (2) minimising the sum of the distances to the line from the misclassified data points.

\begin{figure}
   \includegraphics[width=200bp, bb=10 10 500 700]{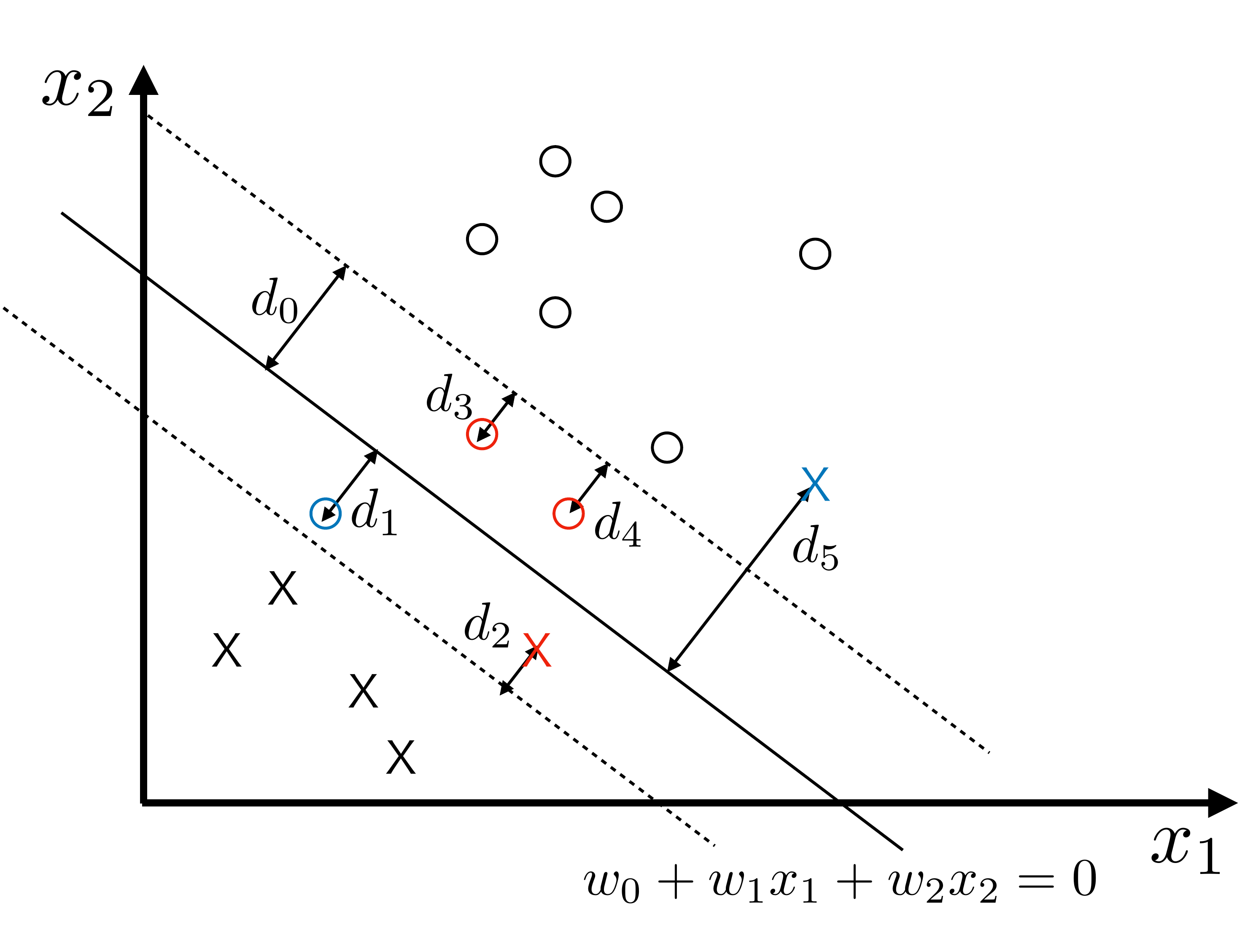}
  \caption{
	The schematic view of LSVC in our case.
  }
  \label{svc}
\end{figure}

Here, we describe the meaning of the second term in more detail.
Figure \ref{svc} displays the schematic view of our analysis.
Suppose that we aim to divide open circles and crosses according to two variables, $x_{1}$ and $x_{2}$.
On the basis of the present $\vb*{w}$, we can calculate $d_{0}$ defined as
\begin{eqnarray}
d_{0} = \frac{1}{\sqrt{w_{1}^{2} + w_{2}^{2}}},
\end{eqnarray}
and select the correctly classified ``near'' data points whose distances to the line are smaller than $d_{0}$, i.e., the correctly classified data points between the dashed lines in the figure (coloured in red).
We also select the misclassified data points regardless of their distances to the line (coloured in blue).
Then, Equation (\ref{loss}) should be written as
\begin{eqnarray}
l = \frac{1}{2}\vb*{w}^{T}\vb*{w} + \frac{C}{d_{0}}\{(d_{1} + d_{5} + 2d_{0}) + (d_{2} + d_{3} + d_{4})\}.
\label{tmploss}
\end{eqnarray}
The first parenthesis in Equation (\ref{tmploss}) corresponds to the sum of the distances to the solid line from the misclassified events (and $d_{0}$ multiplied by the number of them).
The second parenthesis sums up the distances to the nearest dashed line from the correctly classified ``near'' data points.
Finally, by solving the minimisation of $l$ for $\vb*{w}$, we obtained the well separable lines shown in Figure \ref{vrmax-len}.

\section*{Declarations}
\subsection*{\textbf{List of abbreviations}}
\begin{itemize}
	\item CME: Coronal Mass Ejection
	\item LASCO: Large Angle and Spectrometric Coronagraph
	\item SMART:  Solar Magnetic Activity Research Telescope 
	\item SDDI:  Solar Dynamics Doppler Imager 
	\item AIA: Atmospheric Imaging Assembly
	\item LOS: line-of-sight	
	\item LSVC: Linear Support Vector Classification
\end{itemize}

\subsection*{\textbf{Ethics approval and consent to participate}}
Not applicable

\subsection*{\textbf{Consent for publication}}
Not applicable

\subsection*{\textbf{Availability of data and materials}}
See https://www.kwasan.kyoto-u.ac.jp/observation/event/sddi-catalogue/

\subsection*{\textbf{Competing interests}}
Not applicable

\subsection*{\textbf{Funding}}
This work was supported by JSPS KAKENHI grant numbers
JP15H05814 (Project for Solar-Terrestrial Environment Prediction, PSTEP),
JP16H03955, and
JP18J23112.
D. S. is supported by Research Fellowships for Young Scientists from the Japan Society for the Promotion of Science.

\subsection*{\textbf{Authors' contributions}}
\begin{itemize}
	\item Daikichi Seki: statistical analysis, structure and strategy of the paper, and writing the paper
	\item Otsuji Kenichi: developing an algorithm to deduce three-dimensional velocity of filaments.
	\item Takako T. Ishii: observation and calibration
	\item Ayumi Asai: structure and strategy of the paper
	\item Kiyoshi Ichimoto: structure and strategy of the paper
\end{itemize}

\subsection*{\textbf{Acknowledgements}}
D.S. thanks anonymous reviewers for their precious comments.
D.S. also thanks GSAIS Empirical Research Group (GERG) for the discussion to enrich the methodology.
The SOHO/LASCO CME Catalog is generated and maintained at the CDAW Data Center by NASA and The Catholic University of America in cooperation with the Naval Research Laboratory.

%
%

 \bibliographystyle{spmpsci}      
\bibliography{seki5}   
\end{document}